\documentstyle{article}

\begin{document}

\title{Positronium-positronium interaction: Resonance, scattering length, 
and Bose-Einstein condensation}

\author{Sadhan K. Adhikari\\
Instituto de F\'{\i}sica Te\'orica, Universidade Estadual
Paulista,\\
01.405-900 S\~ao Paulo, S\~ao Paulo, Brazil\\}

\date{\today}
\maketitle
\begin{abstract}

The low-energy scattering of ortho positronium (Ps) by ortho Ps has been
studied in a full quantum mechanical  coupled-channel approach.
 In the singlet channel (total spin $s_T=0$) we find S-
and P-wave resonances at 3.35 eV (width 0.02 eV) and 5.05 eV (width 0.04
eV), respectively and a binding of 0.43 eV of Ps$_2$. The scattering
length for $s_T=0$
 is 3.95 \AA \hskip 0.2cm and for $s_T=2$ 
is 0.83 \AA . The small $s_T=2$ scattering length makes the spin-polarized
ortho Ps atoms an almost noninteracting ideal gas which may undergo
Bose-Einstein condensation.

Pacs Numbers: 34.85.+x,  03.75.Fi,  82.30.Gg

\end{abstract} 

\newpage


Of the two types of positronium (Ps), the ortho Ps (triplet)
of large life time 142 ns, compared to that of the para Ps (singlet)  of
life time 0.125 ns, has been of great experimental interest. The
preparation of ortho Ps from positron interaction with atoms and molecules
is well under control. One can  study the
interaction of Ps, a matter-antimatter bound state, with different atoms
and molecules.  There has been precise measurement of
total cross section of  ortho-Ps scattering by H$_2$, N$_2$, He, Ne, Ar,
C$_4$H$_{10}$,
and C$_5$H$_{12}$ \cite{1}.  Such processes may take place on the outer
surface of a star and are also of astrophysical interest. There have also
been studies on bound states of Ps with different atoms \cite{2} as well
as of Ps$_2$ molecule \cite{3}.

In addition to the interaction of Ps with matter, the interaction between
two Ps atoms is also of general interest. As the anti-atom of Ps is also a
Ps, one can  study the interaction between
matter-antimatter (Ps-Ps) states, where the constituents (Ps) are also
matter-antimatter (electron-positron) bound states. In the recent
study of hydrogen-antihydrogen interaction, the constituents
are, however, not matter-antimatter bound states \cite{4}.

Apart from the theoretical interest in testing fundamental symmetries,
(charge conjugation, parity, and time reversal) from a study of
Ps-Ps interaction, the Bose-Einstein condensation (BEC)  of a collection
of spin-polarized ortho Ps (Ps$\uparrow$)  seems to be viable in the
laboratory \cite{5,6}.  The critical
temperature
$T_c$ for BEC of  ideal bosons of mass $m$ and density $n$ is
given by $T_c=$ $(2\pi ^2/m)(n/2.65)^{2/3}$ \cite{6a}.  Hence, a small
mass of Ps at a reasonable density should facilitate BEC by leading to a
large
$T_c$.

Recently, there has been experimental realization of BEC in trapped atoms
($^1$H, $^7$Li, $^{23}$Na, $^{87}$Rb, $^4$He*(2$^3$S$_1$)) at temperatures
of few hundred nK \cite{7,8}.  This initiated intense theoretical and
numerical activities on the subject \cite{9}. In actual experiment a trap
was needed to increase the density of atoms to a desirable level. However,
with the trap emerges a new energy scale $\hbar \omega (\sim $ few nK) in
the problem where $\omega $ is the angular frequency of the trap and a new
$T_c = 0.94 \hbar \omega N^{1/3}$ where $N$ is the number of atoms. For
$N\sim 10^6$ this leads to a $T_c$ of few hundred nK for trapped atoms.

The interaction between the bosonic atoms at low temperature is
proportional to the interatomic scattering length $a$ \cite{9}. In most
cases of trapped atoms above $|a|$ is large \cite{7} which implies a
strong interaction and hence a large deviation from an ideal gas.  This
makes a theoretical investigation via a strongly nonlinear Schr\"odinger
equation extremely complicated \cite{9}.

In the total spin $s_T=0$ state, the collision between two unpolarized
ortho Ps may lead to two para Ps and and their quick eventual
annihilation.  The Ps$_2$ molecule exists in this overall singlet state
where both the electron and positron pairs also remain in singlet states.
Hence, in a collection of unpolarized ortho Ps the long-lived ortho Ps
atoms will disastrously transfer to short-lived para Ps, and for an
efficient BEC one should use  Ps$\uparrow$ \cite{5,6}.  For two
Ps$\uparrow$, $s_T=2$ and both the electron and positron pairs remain in
triplet states.

The small mass of Ps should enhance BEC, and $10^5$ Ps atoms in a
cylindrical cavity in Si of volume $10^{-13}$ cm$^3$ should allow BEC at
tens of
Kelvin temperature according to  the ideal gas BEC formula for $T_c$
\cite{5,6}, in contrast to hundreds of nK for trapped gas
\cite{7,8}. The advantages of a Si cavity over other metalic cavities for
cooling Ps atoms and achieve BEC have been discussed in Ref. \cite{5}.  
If $n|a|^3<<1$, the BEC is close to one of an ideal
gas \cite{5}. It is interesting to calculate the relevant scattering
length for Ps-Ps scattering and see if the BEC of Ps$\uparrow$ can be
considered to be ideal or nearly ideal.

In this Letter we present a full quantum mechanical three-Ps-state
coupled-channel method   with Ps(1s)Ps(1s,2s,2p) states for the study of     
scattering of ortho Ps by ortho Ps 
in both $s_T=0$ (singlet) and $s_T=2$
states. 
We do not consider the simultaneous excitation of two
Ps atoms in this study.
The  $s_T=0$ state  is interesting for studying the Ps$_2$ molecule and
the  $s_T=2$ state
for studying the interaction among the Ps$\uparrow$ atoms to be used
in  BEC. 
Similar coupled-channel approach  was successfully used in Ps
scattering by H
\cite{10}, He \cite{11}, Ar, Ne \cite{12}, Li, Na, K, Rb, Cs \cite{13},
and H$_2$ \cite{14}. In the Ps-H system we could reproduce the
variationally known binding and resonance energies and scattering length
in the singlet channel. In Ps-Li, Ps-Na, Ps-K we could reproduce the
variationally known binding energies \cite{2} in the singlet channel and
predict resonances in S, P, and D waves near the lowest Ps excitation
threshold. In Ps-He, Ps-Ar, Ps-Ne, and Ps-H$_2$ we could give a good
account of the experimental total cross section at low and medium
energies. Hence it seems worthwhile to test this approach in
Ps-Ps scattering.

Here we calculate the Ps-Ps scattering lengths, low-energy phase shifts
and elastic and inelastic cross sections in both $s_T=0$ and $s_T=2$
channels. In the singlet channel ($s_T=0$)  we find resonances in S
and P waves below the lowest Ps excitation threshold.

The theory for the coupled-channel method  of Ps scattering
 has appeared in the literature \cite{10,11}. 
The relevant  coupled Lippmann-Schwinger scattering integral equation in
momentum space is given below
 \begin{eqnarray}\label{aa} &{}&f^\pm _{\nu ' \nu}({\bf
k',k})={
B}^\pm _{\nu '\nu}({\bf k ',k}) \nonumber \\ &-&\frac{1}{2\pi^2}\sum_{\mu}
\int \makebox{d}{\bf q}\frac {{ B}^\pm _{\nu' \mu}({\bf
k',q}) f_{\mu \nu}({\bf q,k}) } {
k_{\mu}^2/2
-q^2/2+\makebox{i}0}, \end{eqnarray} 
where 
$f_{\nu'\nu}^\pm$ is the scattering amplitude, and ${ B}_{\nu
'\nu}^\pm $ is
the corresponding Born amplitude for $s_T =0$ and 2 respectively,
$\nu$ and $\nu '$ denote initial and final states of one of the Ps
while the
other being maintained in the 1s state, 
${\bf k}_\mu $ is the on-shell
relative
momentum of Ps-Ps in the channel $\mu$. 
We present all equations in atomic unit (a.u.) where
$\hbar=e=\bar a_0=m=1$,
where
$e$ (m) is the electronic charge (mass) and $\bar a_0$ is the Bohr radius. 
Because of internal 
 symmetry all direct Born amplitudes with the excitation of only one
Ps, as considered in this Letter, are zero and the scattering proceeds
solely through  exchange interaction by
successive exchange of a pair of electrons between two positrons. 

The exchange Born amplitude for Ps-Ps scattering can be easily
derived
as in Ref. \cite{11} and the details will be elaborated elsewhere. 
The  time-reversal symmetric  exchange Born amplitude is given by  
\begin{eqnarray}\label{xx}
 { B}^\pm_{\nu' \nu}({\bf k',k}) &=& \pm \frac{2}{D}\int \chi^*_{\nu '}
({\bf
r}) e^{{-i\bf Q.r}/2}\chi_\nu ({\bf
r})d{\bf r}  \nonumber \\ &\times& \int \chi^*_{1s} ({\bf
r'}) e^{{i\bf Q.r'}/2}\chi_{1s} ({\bf
r'})d{\bf r'},
\end{eqnarray}
with
$
D= (k^2+k'{}{^2})/4+[\beta_{1s}^2+(\beta_\nu ^2+\beta_{\nu '}^2)/2],
$
where 
${\bf  Q = k - k '}$ is the momentum transfer, $\chi({\bf r})$ the Ps wave
function, and $\beta_\nu ^2$ 
and $\beta_{\nu '}^2$ are binding energies of initial and final states of
Ps in a.u., respectively. The exchange amplitude with $+$
sign in Eq. (\ref{xx}) refers to
an attractive interaction and corresponds to overall $s_T=0$ state
mentioned above where  the electron and positron pairs are 
in singlet states. 
The 
exchange amplitude with $-$ sign refers to a repulsive
interaction and corresponds to the overall 
$s_T=2$ state with both  the
positronium pairs  in the triplet state. 
This exchange Born amplitude  for Ps-Ps
scattering is considered to be a symmetrized generalization of the
Ochkur-Rudge exchange 
amplitude  for electron scattering \cite{15}.

After a partial-wave projection of Eq. (\ref{aa}), the
resultant one-dimensional
scattering equations are solved by the method of matrix inversion. Forty
Gauss-Legendre points are used in the discretization of each
momentum-space integral. 
 
First we study the binding of Ps$_2$
using  the present coupled-channel approach  with
Ps(1s)Ps(1s,2s,2p) states.
 In the singlet channel, this approach  leads
to a Ps$_2$ molecule of binding   0.43  eV in excellent agreement with the  
accurate binding   of 
 0.435445 eV \cite{3} obtained by Frolov and Smith, Jr. using  variational
method.

In Fig. 1 (a) we present the Ps-Ps singlet ($s_T=0$)  phase shifts at
different
Ps-Ps  center of mass energies  ($E_{\mbox
{c.m.}} \equiv
13.6k^2$
eV)  in
S, P, and D waves. It should be noted that $E_{\mbox {c.m.}}=5.1$ eV
corresponds
to the first excitation threshold of one of the Ps.  
In Fig. 1 (b)
we present the same for the $s_T=2$  channel. From  Fig. 1 (a) we
find that
there are S- and P-wave resonances in the singlet channel below the lowest
excitation threshold of the Ps  where the phase shifts jump
by $\pi$. The position and the width of the
resonances are obtained by fitting the corresponding cross sections to a
Breit-Wigner form. The present S- and P-wave  resonance  energies are 
3.35 eV
(width 0.02 eV) and 5.05 eV (width 0.04 eV), respectively.
The appearance of these resonances in the 
electronic singlet state of Ps scattering  by atoms with a single active
electron seems to be a general phenomena and similar resonances have been
found in Ps scattering by H \cite{10} and alkali-metal atoms \cite{13}.
The ability of the present approach  to reproduce the precise binding of 
Ps$_2$ as well as to predict 
resonances in S and P waves assures us of  its
realistic nature and the legitimacy to use it in the $s_T=2$ channel,
where one has 
 essentially the same set of coupled equations (\ref{aa}) with a
change of sign of the
exchange Born amplitudes imposed by the condition of antisymmetrization
for identical fermions. 

Next we report the results for scattering lengths. In the singlet channel
$a_0 = 7.46$ { a.u.} = 3.95 \AA, and in the $s_T=2$ channel $a_2 =
1.56$ { a.u.} = 0.83 \AA. The first is in agreement with the
prediction of Platzman and Mills \cite{5} based on qualitative argument:
$a_0 \simeq 3$ \AA, and should be considered to be refinement over this
estimate. In the interaction of two Ps$\uparrow$, only $a_2$ is pertinent
as the two electrons as well as two positrons have their spins aligned
parallel (triplet states) in two Ps$\uparrow$. It is this scattering
length which governs
the BEC of Ps$\uparrow$.

We exhibit in Fig. 2  the present elastic Ps(1s)Ps(1s), and
inelastic 
Ps(1s)Ps(2s) and  Ps(1s)Ps(2p) cross 
sections at different c.m. energies for Ps-Ps scattering in the two spin
states: $s_T=0$ (full line) and 2 (dashed line). The corresponding exchange
Born contribution (dashed-dotted line) to the cross sections is also shown.
 At high energies the cross sections for both
$s_T=0$ and $s_T=2$ states tend to the corresponding first Born results
and hence to each other as the two Born estimates are the same.  
We find from Fig. 2 that 
the energy-dependence of the elastic cross section is  similar to that for Ps
scattering by H \cite{10} and alkali-metal atoms \cite{13}. The cross
section has a monotonic
slow decrease
with increasing energy. In the $s_T =0$ channel it 
also has a local minimum and maximum below the
lowest Ps excitation threshold manifesting the S- and P-wave resonances.

Let us see the consequence of the present calculation on the proposed BEC
of Ps$\uparrow$ \cite{5,6} of 10$^5$ Ps$\uparrow$ atoms in a Si cavity of
macroscopic volume 10$^{-13}$ cm$^3$ leading to a density $n$ of 10$^{18}$
atoms/cm$^{3}$. The interaction between these atoms will be governed by
the scattering length $a_2\sim 0.8$ \AA \hskip .03cm
and $n|a_2|^3$ gives a good
measure of the nonideal nature of the condensate. In this case
$n|a_2|^3\sim 10 ^{-6} <<1$, and the ideal gas BEC critical temperature
formula without trap should hold good.  The smallness of the scattering
length $a_2$ for Ps$\uparrow$  makes it an excellent candidate for
very weakly interacting BEC. We recall that for the experimentally
observed BEC in trapped atoms, the relevant scattering length for
$^{23}$Na is 27.5 \AA, for $^{87}$Rb is 57.7 \AA, for $^{7}$Li is $-14.5$
\AA, for $^4$He*(2$^3$S$_1$) is 16 \AA \cite{7,8}. The larger scattering
lengths make these atoms strongly interacting 
and their condensates
should have larger deviation from ideal-gas BEC. In the
case of
$^{7}$Li the large negative scattering length implies strong attraction
between the atoms which leads to condensate which
even experience collapse as the condensate grows in size \cite{8}. 

Next we study the profile of the BEC of $N$ atoms of Ps$\uparrow$ in a
spherical Si cavity
of radius $R$ in the ground S state described
by the following mean-field Gross-Pitaevskii (GP) equation in atomic units
\cite{9}
\begin{equation} \label{3}
\left[-\frac{1}{4}\frac{d^2\phi(r)}{dr^2}+\frac{Na_2}{2}\left|
\frac{\phi(r)}{r} 
\right|^2\right]  \phi(r)  =\mu \phi(r),
\end{equation}
where $\mu$ is the chemical potential and  the radial  wave function
$\psi(r)=\phi(r)/r$ is normalized
according to $
\int_0^R |\phi(r)|^2 dr=1.
$ As proposed \cite{5} we consider $N=10^5$ in a spherical cavity of
volume 
$10^{-13}$ cm$^3$ which corresponds to $R=5440$ a.u. We solve
Eq. (\ref{3}) variationally subject to the boundary condition $\phi(0)
=\phi(R)=0$ as in a spherical box. In the ideal gas case $a_2=0$ and the
normalized solution of Eq. (\ref{3}) is $\phi(r)= \sqrt{2/R} \sin(\pi
r/R)$ with $\mu= \pi^2/(4R^2)= 8.34\times 10^{-8}$ a.u.   In Fig. 3 we
plot
the variational
solution of Eq. (\ref{3}) for  $N=10^5$ and $R=5440 $ a.u. and compare
with
this ideal-gas result. The variational result for $\mu$ is $4.18\times
10 ^{-6}$ a.u. The increase in the value of $\mu$ compared to the
ideal-gas
case is due to the repulsion
among the atoms, which contributes positively.
In the ideal gas case the center of the spherical cavity  is the region of
maximum density. In case of Ps$\uparrow$ the region of maximum density has
moved away from the central region due to the repulsive interaction
between the Ps$\uparrow$  atoms. However, as the repulsion among the atoms
is weak the wave function of the BEC of Ps$\uparrow$  atoms is 
qualitatively similar to the ideal gas solution. For BEC of trapped
bosons because of the attraction of the trapping potential the center of
the trap is the region of maximum density for both attractive and
repulsive atomic interactions \cite{9}.

In conclusion, we have studied the Ps-Ps scattering in a coupled-channel
framework allowing the excitation of one of the Ps atoms to Ps(2s,2p)
states. We report the results for 
scattering lengths for total spin $s_T=0$ (3.95 \AA)  and $s_T=2$ (0.83
\AA), and Ps$_2$ binding
energy (0.43 eV). 
We also present results for  phase shifts in lower partial waves
and elastic Ps(1s)Ps(1s)  and inelastic Ps(1s)Ps(2s,2p) cross
sections for $s_T=0$ and $s_T=2$. We find resonances in S and P
waves in the singlet
state. 
The consideration of scattering length
shows that Ps$\uparrow$
atoms are extremely weakly interacting and constitute
excellent candidates for nearly ideal-gas BEC in macroscopic Si cavity
without magnetic trap at a relatively high temperature of tens of Kelvin.
By explicitly solving the GP equation we show that the wave function of the 
Ps$\uparrow$ in a spherical cavity of volume $10^{-13}$ cm$^3$ is similar to 
that of  ideal gas atoms in the same cavity.  

The work is supported in part by the CNPq and FAPESP of Brazil.


\newpage

\vskip 0.5cm
Figure Caption

1. S- (full line), P- (dashed line), and D-wave (dashed-dotted line) phase
shifts of Ps-Ps scattering at different
c.m. energies in the (a) $s_T=0$ and (b) $s_T=2$ states.

2. Cross section of Ps-Ps scattering  at different
c.m. energies in the  $s_T=0$ (full line) and  $s_T=2$
(dashed line) states. The corresponding Born contributions (dashed-dotted line)
are also shown.

3. The wave function $\psi(r)=\phi(r)/r$ of the BEC of Ps$\uparrow$ atoms 
(full line) contrasted with the ideal gas solution (dashed line) in
atomic units.

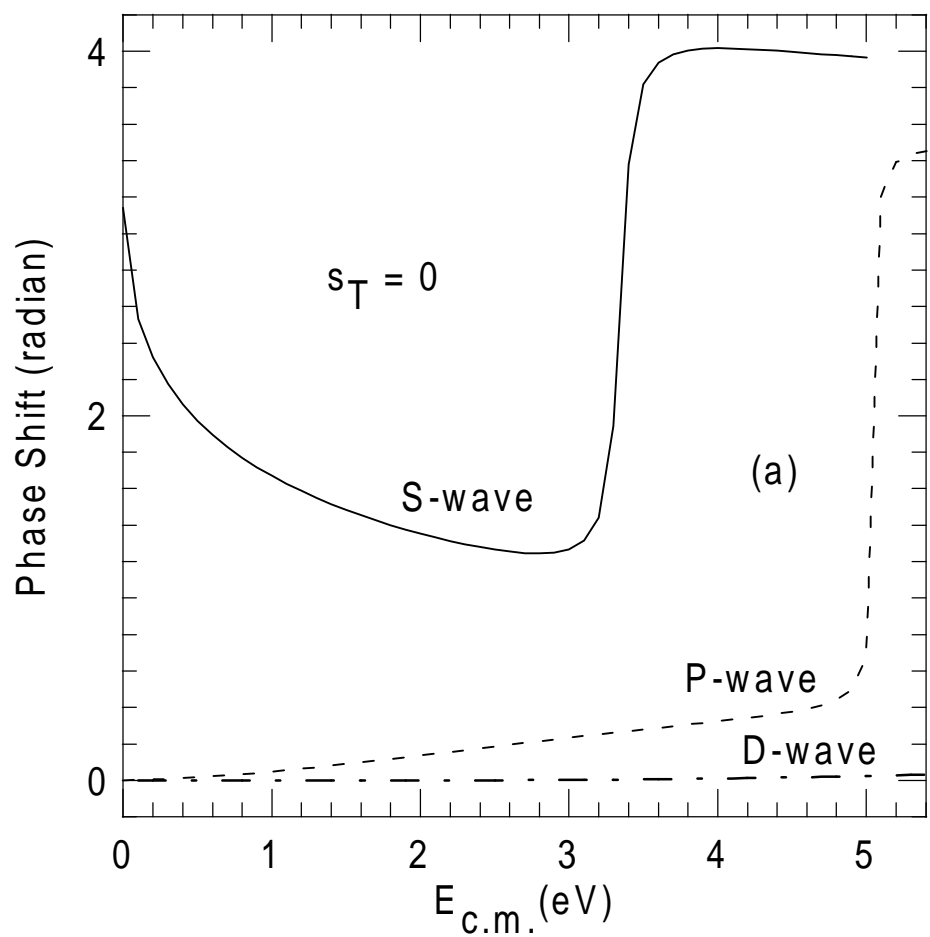
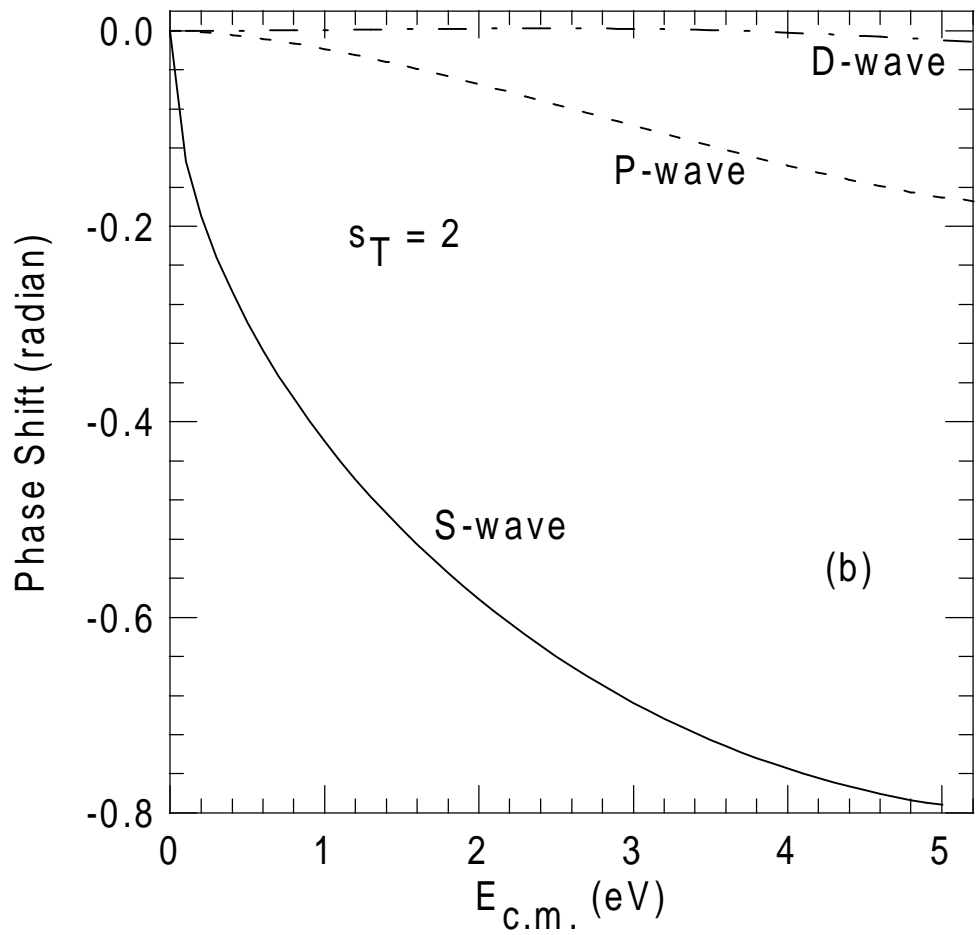
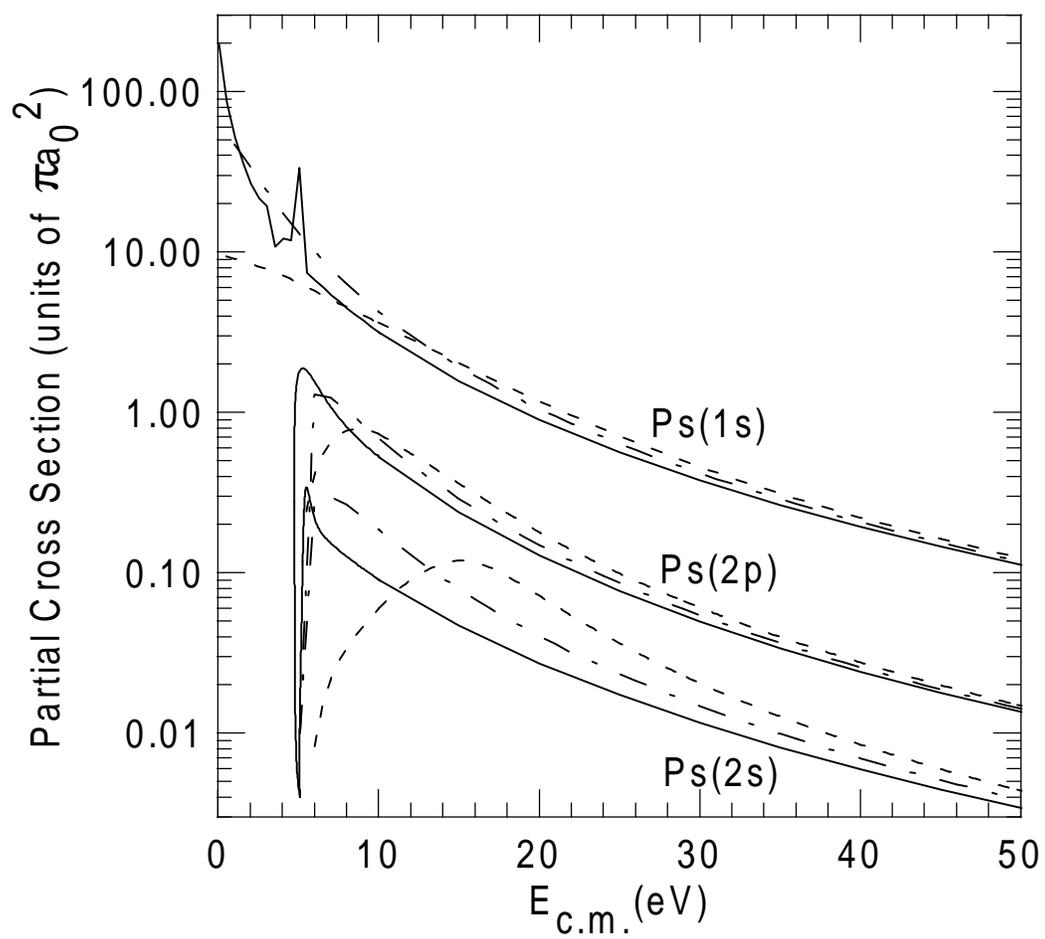
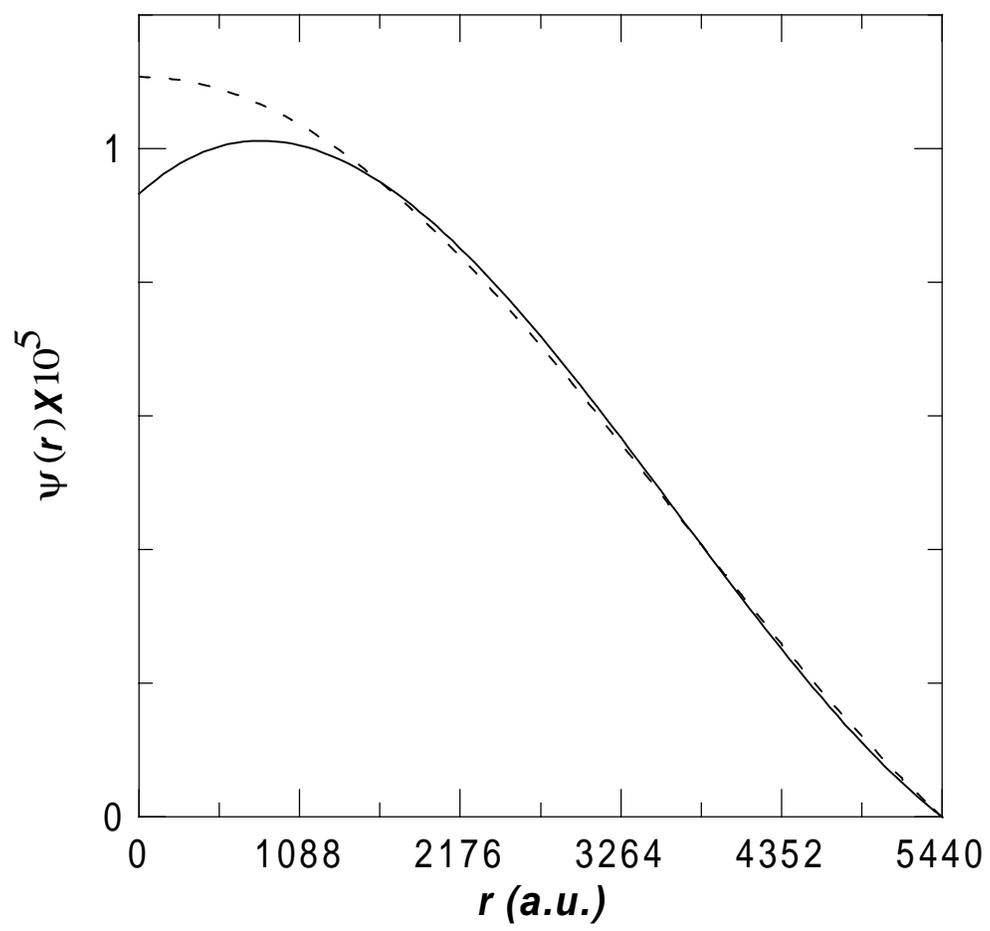
\end{document}